# Two-Dimensional Tail-Biting Convolutional Codes

Liam Alfandary and Dan Raphaeli

School of Elect. Engineering.
Tel. Aviv University.
Tel . Aviv. 69978.
Israel.
Email: liamalfa@eng.tau.ac.il, danr@eng.tau.ac.il

*Abstract* – The multidimensional convolutional codes are an extension of the notion of convolutional codes (CCs) to several dimensions of time. This paper explores the class of two-dimensional convolutional codes (2D CCs) and 2D tail-biting convolutional codes (2D TBCCs ), in particular, from several aspects. First, we derive several basic algebraic properties of these codes, applying algebraic methods in order to find bijective encoders, create parity check matrices and to inverse encoders. Next, we discuss the minimum distance and weight distribution properties of these codes. Extending an existing tree-search algorithm to two dimensions, we apply it to find codes with high minimum distance. Word-error probability asymptotes for sample codes are given and compared with other codes. The results of this approach suggest that 2D TBCCs can perform better than comparable 1D TBCCs or other codes.

We then present several novel iterative suboptimal algorithms for soft decoding 2D CCs, which are based on belief propagation. Two main approaches to decoding are considered. We first focus on a decoder which extends the concept of trellis decoding to two dimensions. Second, we investigate algorithms which use the code's parity check matrices. We apply conventional BP in the parity domain, but improve it with a novel modification. Next, we test the generalized belief propagation (GBP) algorithm. Performance results are presented and compared with optimum decoding techniques and bounds. The results show that our suboptimal algorithms achieve respectable results, in some cases coming as close as 0.2dB from optimal (maximum-likelihood) decoding. However for some of the codes there is still a large gap from the optimal decoder.

I. INTRODUCTION

The class of two-dimensional convolutional codes (2D CCs) and 2D tail-biting convolutional codes (2D TBCCs ), in particular, is a little-researched subject in coding theory. This fact is partly due to the difficulties entailed in representing, encoding and decoding these codes, and partly due to the existence of efficient coding methods (Turbo codes, LDPC codes), which have attracted the interest of researchers in recent years. Nonetheless, 2D CCs are interesting from several aspects. Firstly, these codes generalize the notion of convolutional codes (CCs) in a non-trivial manner, which allows us to broaden the theory behind CCs in general. Secondly, since little research exists in this field, the encoding properties of 2D CCs are largely unknown. Researching these codes will allow us to determine if they have suitable applications, for example, for encoding 2D data (such as images), and as medium size regular error correcting codes.

Multidimensional convolutional codes ($m$-D CCs, where $m$ stands for the dimension) extend the notion of 1D CCs [1] to $m$-D information sequences and generator polynomials. However, in contrast to the one-dimensional case, little research has been done in the field of $m$-D CCs. Significantly, Foransini and Valcher [2] and Weiner [3] have laid out the algebraic theory for $m$-D CCs. Lobo *et al.* [4] have also investigated the subject, concentrating on a sub-family of these codes dubbed "Locally Invertible $m$-D CCs". Charoenlarpnopparut *et al.*, [5] also contributed to the field by suggesting a method for realizing a 2D CC encoder, and constructing its parity check matrix .

Despite the progress made in the field by the above authors and other researchers, the subject of $m$-D CCs still contains many open problems. These problems include: optimal design of $m$-D CCs for specific applications, and the lack of a design procedure; efficient realization of encoders and decoders; and the study of the performance of these codes. Of these problems, two especially stand out: a) What is the encoding power of $m$-D CCs? and b) How to efficiently decode $m$-D CCs? These two problems are the main concern of this paper.

In order to limit the scope of the problem, we chose to focus particularly on 2D tail-biting CCs (2D TBCCs). Tail-biting [1] is a well-known technique to efficiently convert CCs into block codes. By using a cyclic convolution operation rather than the linear convolution used by 'normal' CCs, two benefits are gained: First, the overhead incurred by the linear convolution (the so-called 'tail') is avoided, and thus fewer code bits need to be transmitted for the same coding gain. Second, the cyclic convolution induces invariance to shifts of the codeword, which ensures that all code bits enjoy equal protection (whereas in linear convolution code, bits near the 'tail' enjoy a lesser degree of protection).

Of the papers cited above, none deals directly with $m$-D TBCCs. In this paper we extend the notion of tail biting codes to $m$-D CCs, and derive some algebraic properties for them focusing on rate $1/n$ codes. We show: how to test whether an encoder is bijective; how to test whether it has a polynomial inverse; and how to construct a polynomial inverse matrix and a parity check matrix in case such matrices exist. By construction, 2D TBCCs are cyclic in 2D if every $n$-tuple of outputs is considered as a single code symbol. In addition, it is easy to show that 2D TBCCs are quasi-cyclic in 1D for certain orderings of the 2D sequence in 1D. Then we use a tree-search algorithm to find the distance properties of some sample codes, and compare the implied asymptotic performance with comparable codes.

As for decoding *m*-D CCs, Marrow and Wolf [6] have shown that the existing Viterbi algorithm can be extended to decode two dimensional ISI by transforming the problem to a 1D code, this can be extended to *m*-D CCs. However, the complexity of the algorithm then becomes exponential with the size of the encoded word. This is very undesirable. Weiner [3] and Lobo [4] proposed hard-decision algorithms, for some sub-families of 2D CC but do not give performance results.

In this paper we take a different approach. Like other codes, 2D TBCCs can be described using factor graphs [7], [8]. We discuss two options for creating such a graph. The first construction is carried out in the information sequence domain, and resembles the trellis construction commonly used to decode 1D CCs. Unfortunately, the two dimensional trellis lacks the Markov property essential to the operation of the Viterbi algorithm, thus applying the BP on this graphs leads to a suboptimal algorithm. The second construction is based on the Tanner graph of the code, similar to the decoders used for LDPC codes [9], [10]. LDPC codes are commonly decoded using the loopy belief propagation (LBP) algorithm, which relies on the fact that the Tanner graph of these codes has few short loops. In such a case, the graph is sufficiently close to a tree, so the LBP algorithm effectively approximates the maximum likelihood (ML) decoding algorithm. Contrary to LDPC, the Tanner graph of 2D TBCCs has many short loops. This violates the tree approximation, and therefore the LBP algorithm is not expected to perform close to the ML solution for 2D TBCCs. Nevertheless, we tested the performance of the LBP algorithm on 2D TBCCs, and in some cases it worked surprisingly well. We suggest a modification of the LBP allowing for statistical changing of check equations to improve the convergence of the LBP. Many other cases were decoded more successfully by the more advanced generalized belief propagation (GBP) [11],[12]. The results of the GBP decoder were further improved by a new method of constructing the regions in the GBP. The gap between ML decoding and suboptimal decoding is still too high for some of the codes, suggesting a room for additional decoding algorithm research.

## II. 2D TAIL-BITING CONVOLUTIONAL CODES

A 2D CC of rate $1/n$ operates by convolving a 2D information sequence $u[k_1,k_2]$, where $k_j=0,1,...,N_j-1$, $j=1,2$. with a set of (typically small) 2D sequences $\{g_i[k_1,k_2]\}$, $i=1,2,...,n$, $k_j=0,1,...,K_j-1$, $j=1,2$ called *generator kernels*, to create a set of *n* code sequences,

$$v_i[k_1,k_2] = \sum_{l_1=0}^{K_1-1}\sum_{l_2=0}^{K_2-1} g_i[l_1,l_2]u[l_1-k_1,l_2-k_2] \qquad (1)$$

and the vectors **N**=($N_1,N_2$) and **K**=($K_1,K_2$) are called information support and kernel support respectively.

For finite-support information sequences, the convolution operation has to be terminated. One option is to pad the sequence with zeros along each dimension, which results in $n(N_1+K_1-1)(N_2+K_2-1)-N_1N_2$ additional coded bits, known as *tail bits*. This means that although the nominal code rate is $1/n$, the actual code rate is smaller

$$R_c = \frac{1}{n} \cdot \frac{N_1 N_2}{(N_1 + K_1 - 1)(N_2 + K_2 - 1)} < \frac{1}{n} \qquad (2)$$

An illustration of this loss is shown on the left hand of Figure 1. One way to avoid this rate loss is *tail-biting*, where the linear convolution operation is replaced with a cyclic convolution

$$v_i[k_1, k_2] = \sum_{l_1=0}^{K_1-1} \sum_{l_2=0}^{K_2-1} g_i[l_1, l_2] u[(l_1 - k_1) \bmod N_1, (l_2 - k_2) \bmod N_2] \qquad (3)$$

When using the cyclic convolution in 2D, the edges of the information sequence are "glued" together. This is similar to taking a square piece of paper, and first rolling it lengthwise to a cylinder. Then the ends of the cylinder are brought together to form a *torus*. This geometrical interpretation is shown in Figure 1. This structure implies that when treating each $n$-tuple of outputs, called a *code fragment*, as one symbol, the code is insensitive to shifts in any of the two dimensions.

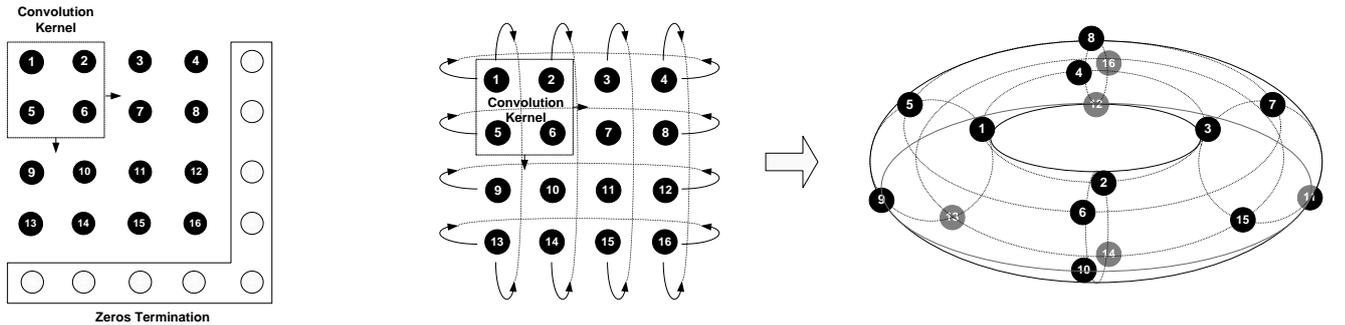

**Figure 1. Convolutional Encoding Process: Zero-Terminated vs. 2D Tail-Biting CC**

III. ALGEBRAIC PROPERTIES

In this section we derive some useful algebraic properties of 2D TBCCs, primarily following [3] and [4]. We focus on rate $1/n$ codes, but many results are easily extended to rate $k/n$ codes.

*Encoder Representation*

A 2D sequence $u[k_1,k_2]$, can be uniquely mapped onto the bivariate polynomial ring $R[x,y]$ via the transform

$$u(x, y) = \sum_{k_1=0}^{N_1-1} \sum_{k_2=0}^{N_2-1} u[k_1, k_2] x^{k_1} y^{k_2} \tag{4}$$

Under this transform, the cyclic convolution operation in eq. (3) becomes

$$v_i(x, y) = g_i(x, y)u(x, y) \bmod I \tag{5}$$

where $I$ is the ideal generated as: $I = \langle x^{N_1-1}, y^{N_2-1} \rangle$.

Thus, all elements of the codeword $\mathbf{v}=(v_0,v_1,...,v_{n-1})$ belong to the quotient ring $R[x,y]/I$. This implies that a 2D TBCC is a *module* of order $n$ over the quotient ring $R[x,y]/I$. We can now use the notation $I$ to denote the ideal as above; $R$ to denote $R[x,y]$; and $\bar{R}$ to denote $R[x,y]/I$, unless noted otherwise.

If we define $\mathbf{G}(x,y) = [g_1(x,y),...,g_n(x,y)]$, then converting eq. (5) to a vector form, we can see that the 2D TBCC encoder induces a map,

$$G(x,y): \bar{R} \to \bar{R}^n \Leftrightarrow \mathbf{v}(x,y) = u(x,y)\cdot \mathbf{G}(x,y) = [u(x,y)\cdot g_1(x,y),...,u(x,y)\cdot g_n(x,y)] \tag{6}$$

For the encoder to be useful, this map has to be *bijective* or *one-to-one*. That is, for any two different elements $u_1, u_2 \in \bar{R}$ the statement $\mathbf{v}_1=u_1\cdot\mathbf{G}(x,y) \neq u_2\cdot\mathbf{G}(x,y)=\mathbf{v}_2$ holds.

Since the code is linear, the above statement is equivalent to requiring $\forall u \in \bar{R}, u \neq 0 \Leftrightarrow \mathbf{v} = u \cdot \mathbf{G} \neq 0$.

**Definition:** A rate $1/n$ encoder $\mathbf{G}(x,y)$ is *non-degenerate* iff there is no input other than $u(x,y)=0$ such that $\mathbf{v}(x,y)=\mathbf{0}$. Alternatively $\mathbf{G}$ is *degenerate* if there exists an input $u(x,y)\neq 0$ such that $\mathbf{v}(x,y)=\mathbf{0}$.

**Proposition:** Let $J$ be the ideal generated by the kernel polynomials $\{g_1(x,y),...,g_n(x,y)\}$. Then a rate $1/n$ tail-biting convolutional encoder $\mathbf{G}(x,y)$ is non-degenerate iff

$$\langle I : J \rangle = I \tag{7}$$

where $\langle I:J \rangle = \{ f \mid \text{for } \forall g \in J, f\cdot g \in I \}$.

**Proof:** The proposition $I \subseteq \langle I:J \rangle$ holds for any two ideals $I$, $J$. On the other hand, if $\mathbf{G}(x,y)$ is degenerate, then there exists $f \notin I$ such that $f \cdot g_j \in I$ for all $j=1,2,...,n$. This means that for this $f$, $f \cdot J \subseteq I$. Therefore, it follows that $f \in \langle I:J \rangle$. Hence, if $\mathbf{G}$ is degenerate, then $\langle I:J \rangle \neq I$. Inverting the statement, if $\langle I:J \rangle = I$ then $\mathbf{G}$ is non-degenerate. ∎

*Parity Check Matrices*

We first consider the concept of parity check matrices, and their application to the decoding of codes.

**Definition:** A parity check matrix $\mathbf{H}(x,y)$ for a generator matrix $\mathbf{G}(x,y)$ of a code $C$ satisfies the following condition: for any $n$-tuple of polynomials $\mathbf{v}(x,y)=[v_1(x,y),\ldots,v_n(x,y)]$, $\mathbf{v}(x,y)\mathbf{H}^T(x,y)=\mathbf{0}$ iff $\mathbf{v}$ is a valid codeword of $C$.

A corollary of this definition is the well known identity: $\mathbf{G}\cdot\mathbf{H}^T=\mathbf{0}$.

**Proposition:** If $(g_1(x,y),\ldots,g_n(x,y))$ have no common divisors except for units, then $\mathbf{G}(x,y)$ has a parity check matrix $n\times(n-1)$ $\mathbf{H}(x,y)$ of the form,

$$\mathbf{H}(x,y) = \begin{pmatrix} g_2 & -g_1 & 0 & \cdots & 0 \\ g_3 & 0 & -g_1 & & \vdots \\ \vdots & & & \ddots & 0 \\ g_n & 0 & \cdots & 0 & -g_1 \end{pmatrix} \tag{8}$$

**Proof**: Let $u(x,y)\in R/I$ and $v(x,y)=u(x,y)\mathbf{G}(x,y)$. Then by construction of $\mathbf{H}(x,y)$:

$v\cdot\mathbf{H}^T = u\cdot[g_1g_2 - g_2g_1,\ldots,g_1g_n - g_ng_1] = \mathbf{0}.$

On the other hand, let $\mathbf{v}\notin\mathrm{Im}\{\mathbf{G}\}$ and suppose that $v\cdot\mathbf{H}=\mathbf{0}$. Then, for $g_1$, $g_j$, $j=2,\ldots,n$ we have, $v_1 g_j + v_j g_1 = 0$. We can write $v_i$ as $v_i = g_i u + r_i$. Where $r_i$ does not divide $g_i$ then,

$0 = (g_1 u + r_1)g_j + (g_j u + r_j)g_1 = r_1 g_j + r_j g_1 \Rightarrow g_1 = -r_1 g_j / r_j$.

Now, since $g_1$ and $g_j$ are polynomials and $r_j$ does not divide $g_j$, then $r_1/r_j$ must be a polynomial, which means that $g_1|g_j$, which contradicts the assumption that $g_1$ and $g_j$ have no common divisors except for units. Therefore $\mathbf{v}\in\mathrm{Im}\{\mathbf{G}\}$ ∎

A more general procedure for constructing parity check matrices for codes of rate $k/n$ is given in [5], which uses Groebner bases.

*Inverse Encoders*

A decoder can be split conceptually into two parts: (a) a codeword estimator which produces an estimate $\hat{v}$ of the codeword $\mathbf{v}$, followed by (b) a pseudo-inverse encoder that recovers the input estimate $\hat{u}$ from $\hat{v}$.

**Definition:** A matrix $\mathbf{G}^{\#}(x,y)=[q_1(x,y),\ldots,q_n(x,y)]$ in $R$ which satisfies

$$\mathbf{G}(x,y)\mathbf{G}^{\#}(x,y)= x^{\alpha}\cdot y^{\beta} \tag{9}$$

is said to be a pseudo-inverse encoder of $\mathbf{G}(x,y)$.

In the case where $\alpha=\beta=0$, $\mathbf{G}^{\#}$ is a true inverse encoder ($\hat{u}=u$), while for $\alpha,\beta\neq 0$, $\mathbf{G}^{\#}$ is a pseudo-inverse encoder ($\hat{u}=x^{\alpha}y^{\beta}u$). In general, $\mathbf{G}^{\#}$ may consist of rational functions (i.e., quotients of polynomials). While the pseudo-inverse in $R$ can also be used in $\overline{R}$, the converse is not always true. The following proposition sets a criterion to determine whether there exists a polynomial $\mathbf{G}^{\#}$ for a given encoder:

**Proposition:** A rate $1/n$ convolutional encoder has a polynomial pseudo-inverse encoder $\mathbf{G}^{\#}(x,y)$ in $R$ iff the ideal $\langle g_1, g_2, ..., g_n \rangle$ contains the monomial $x^{\alpha} \cdot y^{\beta}$, since this means that there exist polynomials $\{q_i(x,y)\}$, $i=1,...,n$, such that

$$\sum_{i=1}^{n} q_i g_i = x^{\alpha} y^{\beta} \qquad (10)$$

This gives by definition the pseudo-inverse, $\mathbf{G}^{\#}(x,y)=[q_1(x,y), q_2(x,y),..., q_n(x,y)]$. The existence of a monomial in an ideal can be easily checked by constructing the Groebner basis for the ideal and checking if it contains a monomial. While this proof is trivial, the proof of the converse is not, and is given in full in Appendix A. The polynomials $\{q_i\}$ are found as a by-product of the Buchberger algorithm for constructing Groebner bases, given in [6].

From now on we shall refer to an encoder which is polynomially pseudo-invertible in $R$ simply as an *invertible* encoder.

## IV. CODE SEARCH

The BEAST algorithm (bidirectional efficient algorithm for searching trees) was formulated in [14] for 1D CCs, both in their zero-terminated and tail-biting forms. The algorithm searches for all codewords with weight $w$. For this aim, forward and backward trees are constructed for each possible starting (and termination) state of the encoder. The trees are kept small in size, so that only paths up to a certain weight and length are stored. Then the forward and backward trees are searched for path pairs that are compatible, i.e., that the forward and backward trees terminate in the same state; their combined length is equal to the length of the information sequence; and that their combined code weight is equal to the desired weight $w$. By searching for $w=0,1,2,...$ etc., the weight spectrum of the code is achieved. Knowing the weight spectrum allows us to calculate the union bound [1], which is an upper bound on the word error probability of a maximum-likelihood decoder over the AWGN channel, where bipolar modulation is assumed

$$p_W\left(\frac{E_b}{N_0}\right) \leq \sum_{w=d_{min}}^{N_1 N_2 n} A(w) \cdot Q\left(\sqrt{2w\frac{k}{n} \cdot \frac{E_b}{N_0}}\right) \qquad (11)$$

The extension of the algorithm to two dimensions is not trivial. The 1D BEAST constructs a tree of the code which consists of states and branches. In the 2D case, we shall define a concept that binds together the concepts of state and branch.

Let $(g_1, g_2,..., g_n)$ be the kernel polynomials of a 2D TBCC of rate $1/n$. Let the kernels have maximum support $K_1 \times K_2$. Then each code fragment at position $(k_1, k_2)$, $\mathbf{v}[k_1,k_2]$ is uniquely determined by $K_1 \times K_2$ information bits out of the sequence $\mathbf{u}$ which we call a *constraint region*.

**Definition:** A *constraint region* of a 2D CC code with kernels $\{g_i\}$, $i=1,...,n$ of support $K_1 \times K_2$ is a sub-sequence of information bits $r(k_1,k_2)=\{u[l_1,l_2],\ l_j=k_j,\ k_j+1,\ ...,\ k_j+K_j-1 \bmod N_j,\ j=1,2\}$. This sub-sequence uniquely determines the code fragment $\mathbf{v}[k_1,k_2]=\{v_1[k_1,k_2],...,v_n[k_1,k_2]\}$.

Using this definition, we can see that the information sequence is composed of many overlapping constraint regions. Thus, we can construct a "2D trellis" composed of adjacent constraint regions. In the 1D case the trellis has one axis, and the BEAST algorithm can advance either forward or backward. In the 2D case, there are two axes, and consequently four directions: down, up, right and left. The up-down direction is determined by decrementing (up) or incrementing (down) the row index $k_1$. The left-right direction is determined by decrementing (left) or incrementing (right) the column index $k_2$.

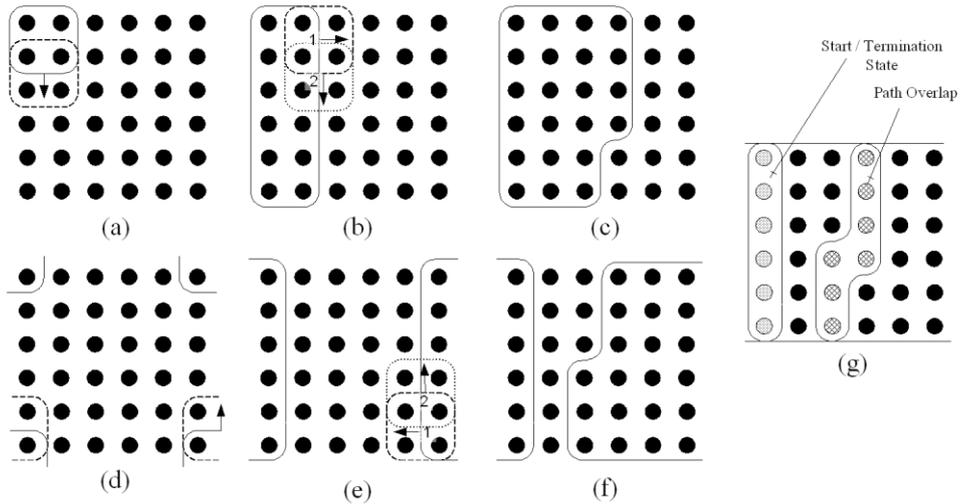

Figure 2. Illustration of the 2D BEAST algorithm

Let us now define compatibility between two constraint regions in a certain direction if they contain the same bits in their overlap region. Here and in the rest of this paper, we will assume modulo $N_j$ in all index calculations.

**Definition:** Let $R_1=\{r_1(k_1,k_2)\}$ and $R_2=\{r_2(k_1,k_2)\}$ be two constraint regions, $k_1=0,...,K_1-1$, $k_2=0,...,K_2-1$. Then $R_2$ is,

a. *Up-compatible* with $R_1$ iff: $r_1(k_1,k_2) = r_2(k_1+1,k_2)$ for $k_1=0,...,K_1-1$, $k_2=0,...K_2-1$
b. *Down-compatible* with $R_1$ iff: $r_1(k_1,k_2) = r_2(k_1-1,k_2)$ for $k_1=0,...,K_1-1$, $k_2=0,...K_2-1$
c. *Right-compatible* with $R_1$ iff: $r_1(k_1,k_2) = r_2(k_1,k_2-1)$ for $k_1=0,...,K_1-1$, $k_2=0,...K_2-1$
d. *Left-compatible* with $R_1$ iff: $r_1(k_1,k_2) = r_2(k_1,k_2+1)$ for $k_1=0,...,K_1-1$, $k_2=0,...K_2-1$

A path in the 2D trellis is defined by a set of adjacent compatible constraint regions. To construct a *forward path* in a 2D trellis is a path we start with the constraint region $r(0,0)$ and proceed in a down-then-right fashion. That is, we advance first in the down direction by laying a sequence of down-compatible regions $r(1,0)$, $r(2,0)$,..., $r(N_1-1,0)$. Once the last row has been reached, we go back to row 0 and advance in the right direction and lay region $r(0,1)$. We then continue in the down direction, laying regions that are down-compatible and right-compatible with previous regions.

To construct a *backward path* in the 2D trellis, we start with the constraint region $r(N_1-1, N_2-1)$ and proceed in the same manner as with the forward path, except that we advance in an up-then-left fashion. That is, we lay up-compatible regions until the first row has been reached. We then wrap around, perform one step left, and continue in the up direction.

The *length* of the path is simply the number of constraint regions it contains. The *weight* of the path is the sum of the weights of all code fragments associated with the path's constraint regions. A pair of forward and backward paths are *compatible* with each other if:

a. The sum of their lengths corresponds to the length of the information sequence.
b. All the constraint regions contained along the boundary of the forward path are compatible with their corresponding regions on the boundary of the backward path (where by boundary of a path we mean the bits at the fringes of the path).

The 2D BEAST algorithm is now defined in the same way as the 1D algorithm, except that states are replaced with constraint regions and that forward and backward paths must now be compatible along their entire mutual borders, rather than terminate in the same state as in the 1D case.

An example of the 2D BEAST algorithm is shown in Figure 2 for a 2D TBCC with kernel support 2x2 over information support 6x6.

Diagrams (a)-(c) show the construction of the forward tree. We start by overlaying the kernel on the top-left corner of the information sequence. We then grow the paths by sliding the kernel down. Since there are four possible compatible regions, our forward tree now contains four possible paths. We continue sliding the kernel down, growing the tree, until we reach the last row of the sequence. We then wrap around to the top (b), and slide the kernel right. If there is still at least one path that has not exceeded the allowed weight, we can then continue sliding the kernel down. In this example, we show a forward path that exceeds the allowed weight and is terminated in (c).

Diagrams (d)-(f) show the construction of the backward tree. This time we start at the bottom-right corner (d) and slide up, until the first row is reached, and then left (c) and up again. The backward path terminates in diagram (f). In Diagram (g) we show the overlapping between the forward and backward paths. The paths have to be compatible, so their overlapping bits are the same. They also have to satisfy the boundary conditions (not shown) imposed by the tail-biting: The leftmost and rightmost columns must be the same as the topmost and bottommost rows. If a pair of paths satisfies all these conditions, we can list its combined weight *w* as a valid codeword of weight *w*.

The problem with this formulation of the algorithm is that the search for compatible pairs of forward-backward paths can be quite cumbersome, since paths are not binned according to their starting state (as opposed to the formulation in the 1D case, where forward and backward paths start and terminate respectively, at the same state). However, the search can be reduced if each path that is constructed is also binned according to its "initial state", which is determined by the $K_2$-1 leftmost columns for a forward column path and the $K_2$-1 rightmost column for a backward path. A pair of compatible forward and backward paths must have the same initial state due to the cyclic convolution. Therefore, we do not need to check for compatibility forward and backward paths which have different initial states, and pairs of potentially compatible paths can be binned according to their initial state in order to speed up the search for compatible pairs.

From a complexity point of view, there is no difference between the 2D and the 1D cases. In each step the tree is appended with two new branches, and then trimmed to a desired size. In fact, the complexity required in order to build the forward and backward trees is about the same if we take 1D TBCC and 2D TBCC and is only marginally dependent on the constraint size.

## V. CODE SEARCH RESULTS

We used the results presented above to search for 2D TBCCs with high minimum distance over various information sequence sizes. As an example, we present results for the search of codes with kernel supports K=2×2 and K=3×3 over information sequences with support N=6×6. The code search was carried out in two parts:

First, a CoCoA [6] program was used to find all non-degenerate codes of rate ½ with a given kernel support. The codes were further categorized as invertible or non-invertible.

Next, the BEAST algorithm was applied to the list of non-degenerate codes in order to find the first few elements of the weight distribution of each code. We use the resulting weight distributions to produce union bound asymptotes for the codes.

In Table I we list the codes with the largest minimum distance found with **K**=(2,2), along with their inverse encoders. Code #1 is considered to be the best of all the codes, since it has the largest minimum distance ($d_{min}$) and the fewest number of words whose weight equals $d_{min}$. This code is also invertible. Code #2 is the second best code, having the same $d_{min}$=6, but with a higher multiplicity. Code #3 is the best systematic code. The union bound asymptotes for these codes are plotted in Figure 3.

In Table II we list the best codes found with K=3×3. Code #4, the best of all the codes, is not invertible. Code #5 is the best invertible code, having the same $d_{min}$ =12, but with a higher multiplicity. And code #6 is the best systematic code. The union bound asymptotes for these codes are plotted in Figure 4. In Table III we give for comparison the weight spectrum of two 1D CCs of rate ½ and $k$=36:

    A. 1D TBCC with generators (13,17). This code has K=4, and $d_{min}$ =6.

    B. 1D TBCC with generators (561,753). This code has K=9, and $d_{min}$ =12.

These codes are the best known 1D CCs under the restrictions of rate and constraint size[1]. Before we compare the spectrums of the 1D TBCCs with the 2D TBCCs, we wish to stress that it is hard to establish a solid basis for this comparison. One possible basis is the memory size, the 1D equivalent to the region that is to be multiplied with the kernel. Using this criterion, code A is to be compared to the 2x2 codes and code B to the 3x3 code. One can see from Table III that the 2D-TBCC codes have fewer words with minimum weight, which translates to better performance. This criterion, however is valid when considering structural properties but not when considering practical decoding complexity. While the 1D codes has decoding algorithm with complexity exponential with the memory size, the decoding complexity of the 2D codes is dependent on the decoding algorithm used, and in the best case (the LBP) is linear in the memory size. When comparing decoding complexities, the LBP decoding of the 3x3 codes is approximately similar to the decoding complexity of code A and much lower than the complexity of code B. Then, even our suboptimal decoding algorithm on code #6 can win the performance of code A with approximately the same complexity. Considering further improvements to the decoding algorithm possible, we believe that the 2D codes will be much better than the 1D codes for same complexity.

The codes are also compared to the sphere packing lower bound (SPLB) of codes with k=36, n=72. This lower bound is given by

$$p_W(n, E_b/N_0) \geq \int_0^\pi \frac{(n-1)(\sin\phi)^{n-2}}{\sqrt{\pi} 2^{n/2} \cdot \Gamma\left(\frac{n+1}{2}\right)} \int_0^\infty s^{n-1} e^{-(s^2+nA^2-2s\sqrt{n}A\cos\phi)/2} ds d\phi \quad (12)$$

where $A = \sqrt{2(k/n)E_b/N_0}$ and $\theta$ solves the equation

$$\Omega_n(\theta) = \int_0^\theta \frac{(n-1)\Gamma\left(\frac{n}{2}+1\right)}{\sqrt{\pi \cdot n} \cdot \Gamma\left(\frac{n+1}{2}\right)} (\sin\phi)^{n-2} d\phi = 2^{-k} \tag{13}$$

Taking Word error rate (WER) of $P_W=10^{-3}$ as the reference point, it can be seen that for K=2×2 the best code is ~1.4 dB and that the best systematic code is ~2.5dB away from SPLB. For K=3×3, the best code found is only ~0.31dB away from the bound, while the best invertible code is ~0.34dB and ~0.35dB away. The best systematic code has $d_{min}=8$, and is ~0.64 dB from the SPLB at $P_w=10^{-3}$. However this gap increases for lower $P_w$ because of its smaller minimum distance.

**TABLE I. BEST CODES WITH K=2×2 OVER N=6×6**

| # | Encoder Kernels | | Inverse Kernels | | $d_{min}$ | $A(d_{min}+i)$ | | | | |
|---|---|---|---|---|---|---|---|---|---|---|
| | $g_1$ | $g_2$ | $q_1$ | $q_2$ | | i=0 | i=1 | i=2 | i=3 | i=4 |
| 1 | 1 1<br>1 0 | 1 1<br>1 1 | 1 | 1 | 6 | 12 | 36 | 72 | 180 | 396 |
| 2 | 1 1<br>1 0 | 1 0<br>1 1 | 1 1 | 1 0 | 6 | 48 | 0 | 306 | 0 | 1440 |
| 3 | 1 0<br>0 0 | 1 1<br>1 1 | 1 | 0 | 5 | 36 | 84 | 72 | 180 | 504 |

TABLE II. BEST CODES WITH K=3×3 OVER N=6×6

| # | Encoder Kernels | | Inverse Kernels | | $d_{min}$ | $A(d_{min}+i)$ | | | | |
|---|---|---|---|---|---|---|---|---|---|---|
| | $g_1$ | $g_2$ | $q_1$ | $q_2$ | | $i=0$ | $i=1$ | $i=2$ | $i=3$ | $i=4$ |
| 4 | 110 110 001 | 101 111 111 | N/A | N/A | 12 | 78 | 1116 | 4158 | 17016 | 60777 |
| 5 | 111 100 010 | 111 111 011 | 101 110 111 111 | 101 110 110 011 | 12 | 276 | 504 | 5382 | 13752 | 66987 |
| 6 | 100 000 000 | 111 110 101 | 1 | 0 | 8 | 36 | 0 | 288 | 0 | 1812 |

TABLE III. BEST 1D CCS WITH COMPARABLE CONSTRAINT SIZE

| Constraint size | Encoder Kernels | | $d_{min}$ | $A(d_{min}+i)$ | | | | |
|---|---|---|---|---|---|---|---|---|
| | $g_1$ | $g_2$ | | $i=0$ | $i=1$ | $i=2$ | $i=3$ | $i=4$ |
| 4 | 13 | 17 | 6 | 36 | 108 | 180 | 396 | 900 |
| 9 | 561 | 753 | 12 | 648 | 0 | 8532 | 0 | 117351 |

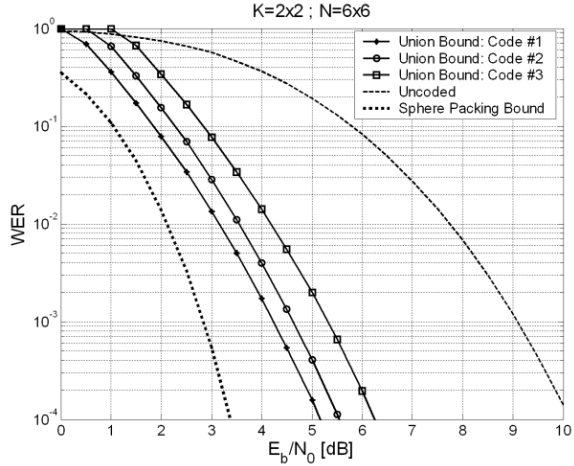 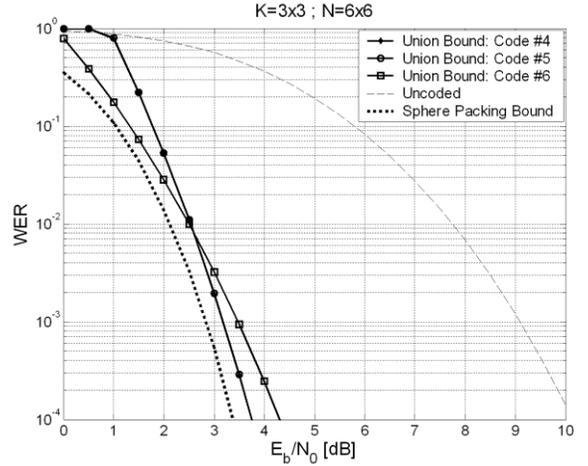

**Figure 3. Union Bound Asymptotes for 2D TBCCs with K=2x2**

**Figure 4. Union Bound Asymptotes for 2D TBCCs with K=3x3**

## VI. INFORMATION PLANE DECODING

In this section we describe decoders that try to directly retrieve the original information word without trying to first reconstruct the codeword.

### A. Optimal Decoding

The problem of optimal decoding of a 2D CC using a maximum-a-posteriori probability (MAP) or maximum likelihood (ML) criterion can be reduced to the problem of decoding a 1D CC over $GF^N(2)$. Optimal decoding (MAP or ML) of a 1D CC memory length $K$ can be achieved with a trellis with $2^{K-1}$ states, irrespective of the input data block length. Both the BCJR and Viterbi algorithms rely on the fact that 1D CC is a causal Markov process, where the conditioning on a single trellis stage effectively separates past and future events.

In the 2D case, there is no immediate notion of causality, and we must condition on a bounding strip (***B***) to create two conditionally independent regions (***A,C***) such that $P(A,C|B) = P(A|B)P(C|B)$. One convenient division is shown in Figure 5.

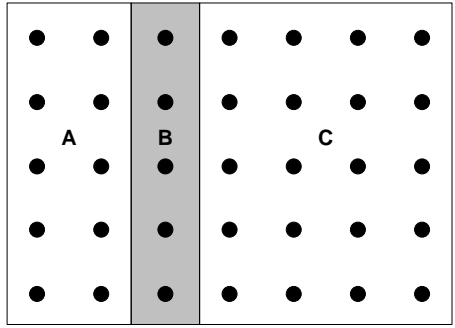

**Figure 5. Division of 2D plane to conditionally independent regions**

Using this division, we can reduce the 2D CC decoding problem to a 1D CC decoding problem by taking each column (row) of the input sequence to be a single input to an equivalent 1D CC encoder, with memory consisting of the $K_2$-1 previous columns. Thus, a single input to the encoder consists of $2^{kN_1}$ bits (where $k$ is the length of the information sequence), and the memory of the encoder consists of $2^{k(K_2-1)N_1}$ bits. The overall complexity of processing a trellis stage, given by the number of states multiplied by the number of branches using this division is, therefore, $2^{kK_2N_1}$. We can choose to decode along columns or rows in order to minimize the trellis complexity.

Since the code is tail-biting, in order to achieve true optimal decoding, the Viterbi algorithm has to be activated $2^{k(K_2-1)N_1}$ times, once for each possible starting / termination state. Then the optimal codeword is selected from the $2^{k(K_2-1)N_1}$ outputs. All this makes optimal decoding very complex and inapplicable in practice except for very short codes. Thus, we must look for sub-optimal algorithms that yield an acceptable performance.

*B. 2D-Trellis Decoder*

The BCJR and max-log-map algorithms were shown in [7], [8] to be respectively equivalent to the sum-product and minimum-sum variants of the loopy belief propagation (LBP) algorithm. This equivalence usually shown by constructing a factor graph where variable nodes are assigned for each encoder state, the information bits and the received channel coded bits, all of which are connected together by factor nodes that represent the trellis structure.

We can construct a similar graph for a 2D TBCC, and apply belief propagation algorithms to it. The resulting algorithm is similar to the BCJR algorithm. However, where the BCJR algorithm messages are passed forward and backward (hence it is also known as the *forward-backward* algorithm), messages in the 2D case are passed *left, right, up* and *down*. However, the performance and even the convergence of the LBP algorithm over this graph are not guaranteed but depend on the properties of the code. In the 1D case the forward-backward algorithm was shown to converge to the MAP solution in the case of tail-biting code even though there is one loop in the graph[7]. For the 2D case, the graph is dense with short loops, so the algorithm is heuristic and does not guarantee convergence to the MAP solution. In fact, for some cases the algorithm fails completely, as will be explained below.

We define the code compatibility function as

$$C(r,v) = \begin{cases} 1 & \text{if } v \text{ is associated with } r \\ 0 & \text{otherwise} \end{cases} \tag{14}$$

where $v \in [0,1]^n$, $r \in [0,1]^{K_1 \times K_2}$. Each region $r(k_1,k_2)$ has four immediate neighbors: $r(k_1+1, k_2+1)$, $r(k_1-1, k_2)$, $r(k_1, k_2+1)$, $r(k_1, k_2-1)$.

We also define horizontal and vertical *intersection compatibility functions* between two neighboring regions as

$$I_V\big(r_1(k_1,k_2), r_2(k_1+1,k_2)\big) = \begin{cases} 1 & \text{if } r_1(k_1,k_2) \text{ is down-compatible with } r_2(k_1+1,k_2) \\ 0 & \text{otherwise} \end{cases} \tag{15}$$

And,

$$I_H\big(r_1(k_1,k_2), r_2(k_1,k_2+1)\big) = \begin{cases} 1 & \text{if } r_1(k_1,k_2) \text{ is right-compatible with } r_2(k_1,k_2+1) \\ 0 & \text{otherwise} \end{cases} \tag{16}$$

where down- and right-compatibility are defined as in Section IV. We use the shorthand notation $C(k_1,k_2)$, $I_H(k_1,k_2)$, $I_V(k_1,k_2)$ to denote the functions that operate specifically on $r(k_1,k_2)$. A section of the resulting graph is shown in Figure 6.

Using the region graph construction from the previous section, we can formulate a message-passing algorithm based on the BP equations. This algorithm is analogous to the 1D BCJR algorithms (using sum-product message passing), but with the following notable differences. First, the algorithm has messages propagating in four directions – up, down, left and right. We assume the receiver is given a set of corrupted code bits, $y_{k_1,k_2} = v_{k_1,k_2} + z_{k_1,k_2}$ where $z_{k_1,k_2}$ are noise samples (e.g., AWGN). We now construct the update rules according to the factor graph message passing equations [8], which are summarized in Table IV.

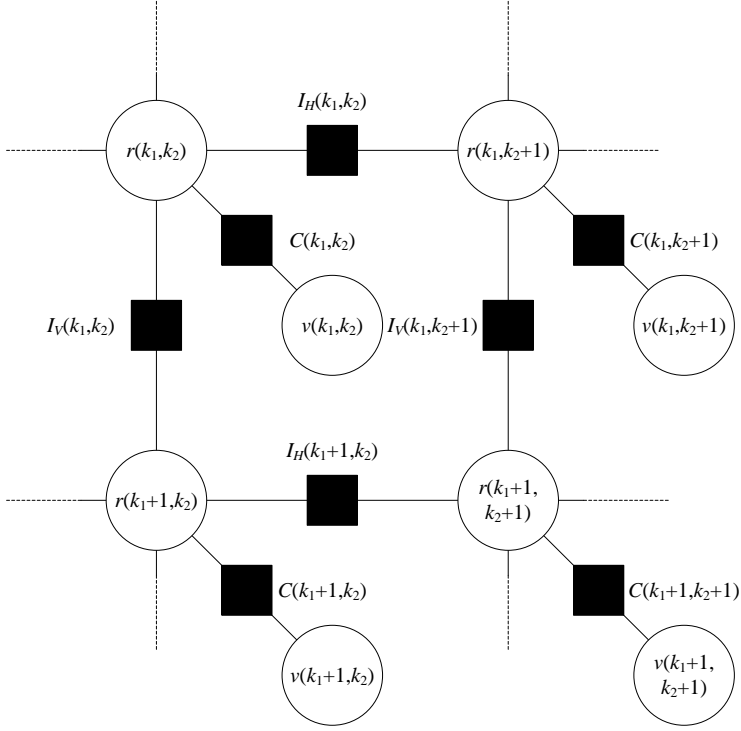

**Figure 6. Constraint region factor graph**

**Table IV. 2D Trellis Message Passing Equations**

| From | To | Message |
|---|---|---|
| $C(k_1,k_2)$ | $R(k_1,k_2)$ | $\lambda^{(0)}_{k_1,k_2}(r) = \alpha \sum_v C(r,v) \cdot p(v \mid y_{k_1,k_2})$ |
| $R(k_1,k_2)$ | $I_H(k_1,k_2)$ | $\alpha^{(l)}_{k_1,k_2}(r) = \lambda_{k_1,k_2} \cdot A^{(l-1)}_{k_1,k_2-1} \cdot H^{(l-1)}_{k_1-1,k_2} \cdot Z^{(l-1)}_{k_1,k_2}$ |
| | $I_H(k_1-1,k_2)$ | $\beta^{(l)}_{k_1,k_2}(r) = \lambda_{k_1,k_2} \cdot B^{(l-1)}_{k_1,k_2} \cdot H^{(l-1)}_{k_1-1,k_2} \cdot Z^{(l-1)}_{k_1,k_2}$ |
| | $I_V(k_1,k_2)$ | $\eta^{(l)}_{k_1,k_2}(r) = \lambda_{k_1,k_2} \cdot A^{(l-1)}_{k_1,k_2-1} \cdot B^{(l-1)}_{k_1,k_2} \cdot H^{(l-1)}_{k_1,-1,k_2}$ |
| | $I_V(k_1,k_2-1)$ | $\zeta^{(l)}_{k_1,k_2}(r) = \lambda_{k_1,k_2} \cdot A^{(l-1)}_{k_1,k_2-1} \cdot B^{(l-1)}_{k_1,k_2} \cdot Z^{(l-1)}_{k_1,k_2}$ |
| $I_H(k_1,k_2)$ | $R(k_1+1,k_2)$ | $A^{(l)}_{k_1,k_2}(r) = \sum_{r'} I_H(r,r') \cdot \alpha^{(l)}_{i,j}(r')$ |
| | $R(k_1,k_2)$ | $B^{(l)}_{k_1,k_2}(r) = \sum_{r'} I_H(r',r) \cdot \beta^{(l)}_{k_1,k_2}(r')$ |
| $I_V(k_1,k_2)$ | $R(k_1,k_2+1)$ | $H^{(l)}_{k_1,k_2}(r) = \sum_{r'} I_V(r,r') \cdot \eta^{(l)}_{k_1,k_2}(r')$ |
| | $R(k_1,k_2)$ | $Z^{(l)}_{k_1,k_2}(r) = \sum_{r'} I_V(r',r) \cdot \zeta^{(l)}_{k_1,k_2}(r')$ |

The 2D Trellis algorithm has a complexity that is linear with the input dimensions, but exponential with the support of the kernel. Thus the number of operations per bit is of order $O(2^{K_1 K_2})$. This is because the graph has $N_1 N_2$ region nodes, and each of them has to compute $2^{K_1 K_2}$ beliefs. The actual number of operations can be reduced by a sophisticated implementation.

We now formulate a necessary (but not sufficient) condition for the convergence of the algorithm. Let us begin by analyzing the relationship between a constraint region and the code fragment associated with it. First, we break the region into two sub-regions: The intersection with a neighboring region (for example on the left)

$$S^L_{i,j} = R_{i,j} \cap R_{i,j-1} \quad ; \quad T^R_{i,j-1} = R_{i,j-1} \setminus S^L_{i,j} \tag{17}$$

The sub-regions denoted by *S* correspond to the notion of a *state* in the 1D trellis; those denoted by *T* correspond to the notion of a *branch* in the 1D trellis. Using this notation we can re-write the message-passing equations of the intersection nodes as,

$$A_{i,j}^{(l)}(s_L, t_R) = \sum_{t'_L} \left[ p(t'_L, s_L \mid y_{i,j-1}) \cdot A_{i,j-1}^{(l-1)}(t'_L, s_L) \cdot H_{i-1,j}^{(l-1)}(t'_L, s_L) \cdot Z_{i,j}^{(l-1)}(t'_L, s_L) \right] \qquad (18)$$

The code fragment may now be represented as a combination of the sub-regions **S** and **T**

$$v_i = \sum_{k_1=0}^{K_1-1} \sum_{k_2=0}^{K_2-2} s(k_1, k_2) \cdot g_i(K_1 - k_1 - 1, K_2 - k_2 - 1) + \sum_{k_1=0}^{K_1-1} t(k_1, 0) \cdot g_i(K_1 - k_1 - 1, K_2 - k_2 - 1) \qquad (19)$$

for $i=1,2,...,n$. We can write this in shorthand vector form

$$\mathbf{v} = [\mathbf{g}'_0, ..., \mathbf{g}'_{n-1}]^T \mathbf{s} + [\mathbf{g}''_0, ..., \mathbf{g}''_{n-1}]^T \mathbf{t} = \mathbf{G}' \cdot \mathbf{s} + \mathbf{G}'' \cdot \mathbf{t} \qquad (20)$$

where **s** is the column stack of the sub-region *S*, **t** is the column stack of sub-region *T*, and $\{\mathbf{g}'\}$ and $\{\mathbf{g}''\}$ are column vectors corresponding to the relevant elements of the kernels $\{\mathbf{g}\}$.

For the algorithm to work, we want to know that only some of the possible sub-regions (not all of them) are mapped to a single code fragment. Since the code is linear, we only need to check that not all possible sub-regions **T** or **S** give **c=0**. Assuming that **c=0**, then

$$\mathbf{G}' \cdot \mathbf{s} = -\mathbf{G}'' \mathbf{t}. \qquad (21)$$

We have to impose some constraints on $\mathbf{G}'$ and $\mathbf{G}''$ to ensure good message passing.

**Proposal (Necessary condition for 2D message passing):** Non-trivial message passing occurs in a horizontal (vertical) direction if

$$\text{Im}\{\mathbf{G}'\} \neq \text{Im}\{\mathbf{G}\}. \qquad (22)$$

**Proof:** Consider messages passed in a certain direction, for example, to the right. Assume that $\text{Im}\{\mathbf{G}'\} = \text{Im}\{\mathbf{G}\}$. Then for every **s** there exists a unique **t** such that $\mathbf{G}'\mathbf{s} + \mathbf{G}''\mathbf{t} = 0$. Now let us look at the first iteration of the algorithm, assuming the all-zero codeword has been transmitted and received correctly. The region beliefs are initialized to

$$P(\mathbf{r}|\mathbf{c} = \mathbf{0}) = \begin{cases} 1/|\text{Im}\{\mathbf{G}\}| & \text{if } \mathbf{G} \cdot \mathbf{r} = \mathbf{0} \\ 0 & \text{otherwise} \end{cases}$$

The message passed in the right direction is therefore

$$A_{i,j}^{(l)}(s_L, t_R) \propto \sum_{t'_L} p(r|c=0) = \sum_{t'_L} p(t'_L, s_L | c=0) = p(s_L | c=0) = 1/|\text{Im}\{\mathbf{G}\}|$$

where the last equation follows from the fact that only a unique **t** satisfies **Gr=0** with any **s**, so any non-zero value of $P(\mathbf{r}|\mathbf{c=0})$ is summed only once. Since the code is linear, the same holds also for **c≠0**. This means that the messages passed in the right direction are constant, which means trivial message passing.

For the algorithm to perform well <u>we require that non-trivial message passing shall occur in at least one horizontal (left or right) direction and one vertical (up or down).</u>

## VII. PARITY-CHECK DECODERS

An alternative approach for soft decoding is to try to estimate the transmitted codeword, and then apply an inverse encoding operation to recover the original information word. This is depicted in Figure 7. We refer to these algorithms as *BP in the parity check domain*.

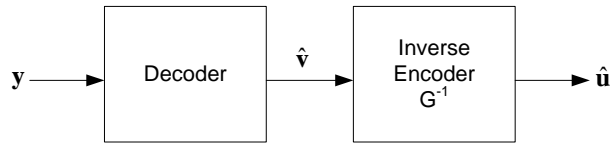

**Figure 7. Two Stage Decoding: Decoder and Inverse Encoder**

For systematic codes, the original information is part of the codeword itself and can be easily extracted. For non-systematic codes, a non-trivial inverse encoding has to be performed. Unfortunately, in the case of a decoding error, the inverse encoding operation magnifies the number of errors such that a single erroneous code bit may cause many erroneous information bits. Decoding in the parity check domain requires the use of a parity check matrix or *syndrome* formers. We limit ourselves to discussing codes of rate $1/n$, for which the syndrome formers are readily available.

### A. Classical Loopy Belief Propagation

In this approach we construct a classical bipartite Tanner graph [10] normally used for decoding LDPC. This bi-partite graph is based on the parity check matrix of the code. Given the following rate $1/n$ code: $\mathbf{G}(x,y)=[g_1(x,y), g_2(x,y),\ldots, g_n(x,y)]$, one possible parity check matrix is given by equation (8). An example is shown in Figure 9. Each square in the figure denotes a check equation and a circle denote the coded bits, empty and full for the two output bits, respectively.

BP in the parity check domain is advantageous in several aspects. First, it converts the problem of decoding a 2D CC into the known Gallager BP algorithm [9], widely used to decode LDPC codes. Next, the fact that messages are scalar LLRs instead of belief vectors reduces the amount of computations and memory required to store the messages. Finally, due to the regular structure of the graph, this decoder can be easily implemented in hardware. However, the graph constructed in this manner has many loops, and is not guaranteed to converge to the MAP/ML solution for any code.

We further wish to integrate the inverse decoder into the decoder as mentioned above to reconstruct the information sequence $u$ from the coded bits $\{v_i\}$. For this purpose we create a Tanner graph of the inverse encoder matrix, and combine it with the code graph, as follows.

Let us assume that the encoder has a pseudo-inverse $\mathbf{G}^{\#}(x,y)$, with delay $x^\alpha \cdot y^\beta$, such that $x^\alpha \cdot y^\beta \mathbf{u} = \mathbf{G}^{\#}\mathbf{v}$. Then we can write the following equation,

$$\mathbf{G}^{\#}\mathbf{v} + x^\alpha y^\beta \mathbf{u} = 0 \tag{23}$$

This equation can be treated as an additional check equation and added to the Tanner graph. Note that the information nodes $u$ can be looked at as if they were transmitted but then erased (i.e., have an initial LLR of zero). Since the erased nodes do not contribute anything to the decoding process, we do not have to incorporate equation (23) directly to the decoding network. Rather, it is more useful to use a decoding network based only on $\mathbf{H}$, and construct a separate inverter network based on $\mathbf{G}^{\#}$. The decoding network $\mathbf{H}$ receives corrupted LLRs at its input, and produces improved LLRs at its output. These improved LLRs are fed to the inverter network $\mathbf{G}^{\#}$, where single belief propagation iteration acts as an optimal "soft combiner" of code bit LLRs, to produce information bit LLRs.

From a complexity point of view, the LBP algorithm has a relatively low computational complexity. The number of operations is linear with the dimensions of the input, and polynomial with weight of the kernel matrix. Let $wt(g) = \max_{i=1...n}\{wt(g_1) + wt(g_i)\}$ be the combined weight of the generator polynomials. The graph has $(n-k)N_1 N_2$ check nodes and $nN_1 N_2$ variable nodes. The complexity is governed by the check nodes, with each check node connected to $wt(g)$ variable nodes. Therefore a check node is required to compute $wt(g)[wt(g)-1]$ messages at each iteration. Thus the complexity per information bit is $O(wt^2(g))$. Note that by construction $wt(g) \leq nK_1 K_2$.

## B. Modified Loopy Belief Propagation

The problem with the LBP is that in many cases the algorithm does not converge into a legal codeword. Therefore, we modified the algorithm to continue its search; relying on the fact that more than one syndrome former exists for each code. Let $z(x,y) \in R/I$ be an arbitrary polynomial then

$$\mathbf{H}(x, y) = \begin{pmatrix} g_2 z & -g_1 z & 0 & \cdots & 0 \\ g_3 z & 0 & -g_1 z & & \vdots \\ \vdots & & & \ddots & 0 \\ g_n z & 0 & \cdots & 0 & -g_1 z \end{pmatrix}$$

is a syndrome former. Thus, we can easily create few syndrome formers $\mathbf{H}_k(x, y)$, $k = 1 \ldots T$ by convolving the original kernels with some small matrices $z_k(x, y)$ such that the weight of the result will be low (see Example 1). Then, we run the loopy BP with one change (see pseudo code below): if after some constant number of iterations the algorithm has not converged into a legal codeword, then we randomly replace a fraction of the rows of the current syndrome former with rows taken from one of $\mathbf{H}_k(x, y)$. We then let the algorithm continue with this new syndrome former, by initializing the messages related to the non-deleted rows with the messages from the previous iteration and initializing the new rows with the original LLR's received from the channel. If after a fixed number of steps the algorithm has not converged, we restart the algorithm with another randomly chosen syndrome former and the original LLR's. This random matrix is generated by mixing rows from all $\mathbf{H}_k(x, y)$. The modified LBP has often resulted in a much better performance compared to the LBP. As from a complexity point of view, the complexity is about the same as the LBP, and the average number of iterations is close to 1 in high SNR.

This method of improving the LBP cannot be used for standard LDPC unless there is a method to create few parity check matrices that are as sparse as the original one.

**Example 1:** Let us consider the following 2×2 code over 6×6 sequences

$$G(x, y) = (1 + x + y, 1 + x + y + xy) \Leftrightarrow g_1 = \begin{pmatrix} 1 & 1 \\ 1 & 0 \end{pmatrix} g_2 = \begin{pmatrix} 1 & 1 \\ 1 & 1 \end{pmatrix}$$

The syndrome former is

$$H(x, y) = (1 + x + y + xy, 1 + x + y) \Leftrightarrow h_1 = \begin{pmatrix} 1 & 1 \\ 1 & 1 \end{pmatrix} h_2 = \begin{pmatrix} 1 & 1 \\ 1 & 0 \end{pmatrix}$$

We choose $\{z_k\}$ to be those generating the three lowest weight error pattern, weights 7,8,8 (see below).

$$z_1(x,y)=(1), z_2(x,y)=(1+y), z_3(x,y)=(1+x) \Leftrightarrow z_1 = \begin{pmatrix} 1 & 0 \\ 0 & 0 \end{pmatrix}, z_2 = \begin{pmatrix} 1 & 0 \\ 1 & 0 \end{pmatrix}, z_3 = \begin{pmatrix} 1 & 1 \\ 0 & 0 \end{pmatrix}$$

Thus we have $\{H_k\}_{k=1}^{3}$

$$H_1(x,y) = (1+x+y+xy, 1+x+y) \Leftrightarrow h_1 = \begin{pmatrix} 1 & 1 \\ 1 & 1 \end{pmatrix} h_2 = \begin{pmatrix} 1 & 1 \\ 1 & 0 \end{pmatrix}$$

$$H_2(x,y) = (1+x+y^2+xy^2, 1+x+xy+y^2) \Leftrightarrow h_1 = \begin{pmatrix} 1 & 1 & 0 \\ 0 & 0 & 0 \\ 1 & 1 & 0 \end{pmatrix} h_2 = \begin{pmatrix} 1 & 1 & 0 \\ 0 & 1 & 0 \\ 1 & 0 & 0 \end{pmatrix}$$

$$H_3(x,y) = (1+y+x^2+x^2y, 1+y+xy+x^2) \Leftrightarrow h_1 = \begin{pmatrix} 1 & 0 & 1 \\ 1 & 0 & 1 \\ 0 & 0 & 0 \end{pmatrix} h_2 = \begin{pmatrix} 1 & 0 & 1 \\ 1 & 1 & 0 \\ 0 & 0 & 0 \end{pmatrix}$$

**Pseudo-code for Modified LBP:**

**Input:** LLR from the channel

- Let $R = N_1 N_2$, $C = N_1 N_2 n$, i.e. the size of the resulting parity check matrix
- Create $H_{i,j}^k$, the matrix version of $\{\mathbf{H}_k(x,y)\}_{k=1}^T$, where $i = 0 \ldots R-1$, $j = 0 \ldots C-1$
- For $p=1:P$ (number of outer iterations)
    - $\theta$ = random $T$-ary vector of length $R$
    - $\hat{H}_{i,j} = H_{i,j}^{\theta_i}$, $i = 0 \ldots R-1$, $j = 0 \ldots C-1$
    - $[c \; V] = LBP\{LLR, \hat{H}\}$ where c is the decoded output (at the current iteration) and $V$ is a matrix of messages from variables to checks: $v_{i,j} = v_{i \to j}$
    - For $q=1:Q$ (number of inner iterations)
        - If $c \cdot (H^1)^T = 0$
            - Output c
        - Else
            - $\Delta \theta_i = \{1 \text{ in prob. p, 0 in prob. } (1-p)\}$, $i = 0 \ldots R-1$
            - $\theta_i = (\theta_i + \Delta \theta_i) \mod T$
            - $\hat{H}_{i,j} = H_{i,j}^{\theta_i}$, $i = 0 \ldots R-1$, $j = 0 \ldots C-1$
            - For $i=0:R$-1
                - If $\Delta \theta_i = 1$ and for each $j$ s.t. $\hat{H}_{i,j} = 1$
                    - $V_{i,j} = LLR_j$
                - End if
            - End loop
            - $[c \; V] = LBP\{V, \hat{H}\}$
        - End if
    - End inner loop
- End outer loop

## C. Generalized Belief Propagation

Generalized belief propagation [11]-[12] seeks to avoid the problems caused by loops in the graph by grouping nodes together into regions, and passing messages between these regions rather than individual nodes. The first stage of the algorithm is to construct a valid region graph, where a region is some subset of the factor graph containing variables and check nodes. We have followed the Kikuchi approximation construction described in [11], with a modification that will be explained below.

The standard Kikuchi approximation begins by defining the set of large regions in the graph, from which intersections will be constructed. A large region is defined to contain all variable nodes (which correspond to code bits) that are connected to a single factor node (the parity check equation in which the code bits participate). Denote the set of large regions as $\Gamma_1$, and this is layer 1 in the algorithm. The rest of the groups of regions in the next layers are defined recursively. The group of regions $\Gamma_i$ of the $i$'th layer is composed of all the intersections between the regions in $\Gamma_{i-1}$. This process continues until no more new intersections can be found.

We have found that this construction performs very poorly in some cases, and in order to improve robustness we have introduced the following modification. After all the intersections have been found as above, we form more intersections between nodes of different layers, i.e., we allow the formation of new intersections from the entire set of existing regions $\tilde{R} = \Gamma_1 \cup \Gamma_2 \cup ... \cup \Gamma_N$ where $\Gamma_N$ is the last layer. However, we do not allow intersections to form between a region and any of its ancestors.

The next example demonstrates the different constructions on a 2D TBCC.

**Example:** Let us consider the following 2×2 code over 4×4 sequences

$$G(x, y) = (x + y, 1 + x + y) \iff g_1 = \begin{pmatrix} 0 & 1 \\ 1 & 0 \end{pmatrix} g_2 = \begin{pmatrix} 1 & 1 \\ 1 & 0 \end{pmatrix}$$

The syndrome former is

$$H(x, y) = (1 + x + y, x + y) \iff h_1 = \begin{pmatrix} 1 & 1 \\ 1 & 0 \end{pmatrix} h_2 = \begin{pmatrix} 0 & 1 \\ 1 & 0 \end{pmatrix}$$

We number the variable nodes (code bits) of the codeword from 0 to 31. According to the connectivity between variable and check nodes, we can form the following 16 large regions (represented as sets of code bits):

| $R_0=\{0,3,12,19,28\}$ | $R_4=\{0,4,7,16,23\}$ | $R_8=\{4,8,11,20,27\}$ | $R_{12}=\{8,12,15,24,31\}$ |
|---|---|---|---|
| $R_1=\{0,1,13,16,29\}$ | $R_5=\{1,4,5,17,20\}$ | $R_9=\{5,8,9,21,24\}$ | $R_{13}=\{9,12,13,25,28\}$ |
| $R_2=\{1,2,14,17,30\}$ | $R_6=\{2,5,6,18,21\}$ | $R_{10}=\{6,9,10,22,25\}$ | $R_{14}=\{10,13,14,26,29\}$ |
| $R_3=\{2,3,15,18,31\}$ | $R_7=\{3,6,7,19,22\}$ | $R_{11}=\{7,10,11,23,26\}$ | $R_{15}=\{11,14,15,27,30\}$ |

The large regions will form the top layer of the region graph. From these regions we can form the following 16 intersection regions, which will form the middle layer of the region graph:

| $R_{16}=R_0 \cap R_7=\{3,19\}$ | $R_{20}=R_2 \cap R_5=\{1,17\}$ | $R_{24}=R_4 \cap R_{11}=\{7,23\}$ | $R_{28}=R_8 \cap R_{15}=\{11,27\}$ |
|---|---|---|---|
| $R_{17}=R_0 \cap R_{13}=\{12,28\}$ | $R_{21}=R_2 \cap R_{15}=\{14,30\}$ | $R_{25}=R_4 \cap R_8=\{4,20\}$ | $R_{29}= R_9 \cap R_{12}=\{8,24\}$ |
| $R_{18}=R_1 \cap R_4=\{0,16\}$ | $R_{22}=R_3 \cap R_6=\{2,18\}$ | $R_{26}=R_6 \cap R_9=\{5,21\}$ | $R_{30}=R_{10} \cap R_{13}=\{9,25\}$ |
| $R_{19}=R_1 \cap R_{14}=\{13,29\}$ | $R_{23}=R_3 \cap R_{12}=\{15,31\}$ | $R_{27}=R_7 \cap R_{10}=\{6,22\}$ | $R_{31}=R_{11} \cap R_{14}=\{10,26\}$ |

The original construction would terminate here, since there are no new intersections to be found in the last layer! Construction II would continue by attaching a pair of single variable regions to each of the regions found in layer 2.

However, the modified construction goes on to form a third layer in which we have 16 small regions, composed of single variable nodes, and formed by intersections of nodes from layer 1 and layer 2:

| $R_{32}=R_{18} \cap R_0=\{0\}$ | $R_{36}=R_{25} \cap R_5=\{4\}$ | $R_{40}=R_{29} \cap R_8=\{8\}$ | $R_{44}=R_{17} \cap R_{13}=\{12\}$ |
|---|---|---|---|
| $R_{33}=R_{20} \cap R_1=\{1\}$ | $R_{37}=R_{26} \cap R_5=\{5\}$ | $R_{41}=R_{30} \cap R_9=\{9\}$ | $R_{45}=R_{19} \cap R_{13}=\{13\}$ |
| $R_{34}=R_{22} \cap R_2=\{2\}$ | $R_{38}=R_{27} \cap R_6=\{6\}$ | $R_{42}=R_{31} \cap R_{10}=\{10\}$ | $R_{46}=R_{21} \cap R_{14}=\{14\}$ |
| $R_{35}=R_{16} \cap R_3=\{3\}$ | $R_{39}=R_{24} \cap R_7=\{7\}$ | $R_{43}=R_{28} \cap R_{11}=\{11\}$ | $R_{47}=R_{23} \cap R_{15}=\{15\}$ |

Note that there are no single-variable regions that contain variable nodes 16 to 31.

In Figure 8 the nine central largest regions of this graph are shown (the tail biting regions are omitted for clarity). It can be seen that for each region the two bottom left variable nodes coincide with the two top right variable nodes of an adjacent region. These intersections form the middle layer of the graph.

A part of the resulting region graph is shown in Figure 9. The "ancestry" of the variable nodes that are members of the central region is shown. The regions and connections that result from the modified construction are shown in a dashed line.

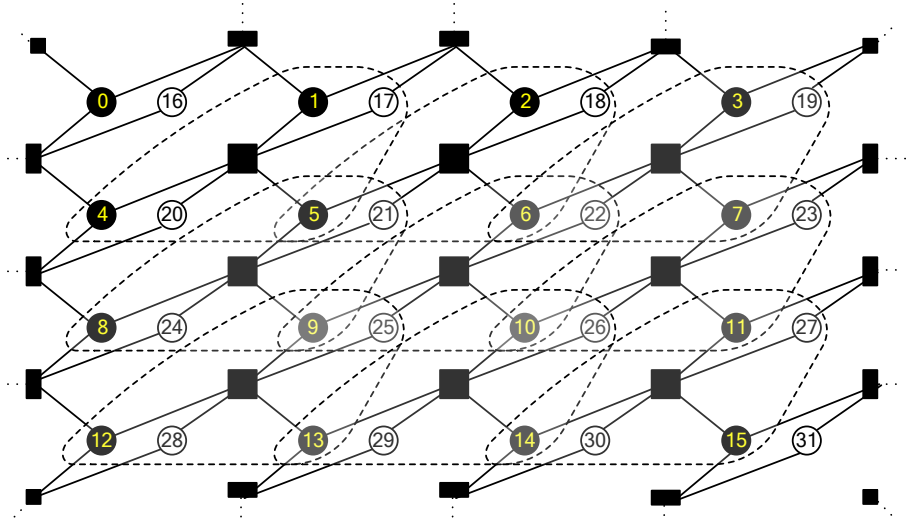

Figure 8. Tanner graph of 2D TBCC with overlaid GBP large regions

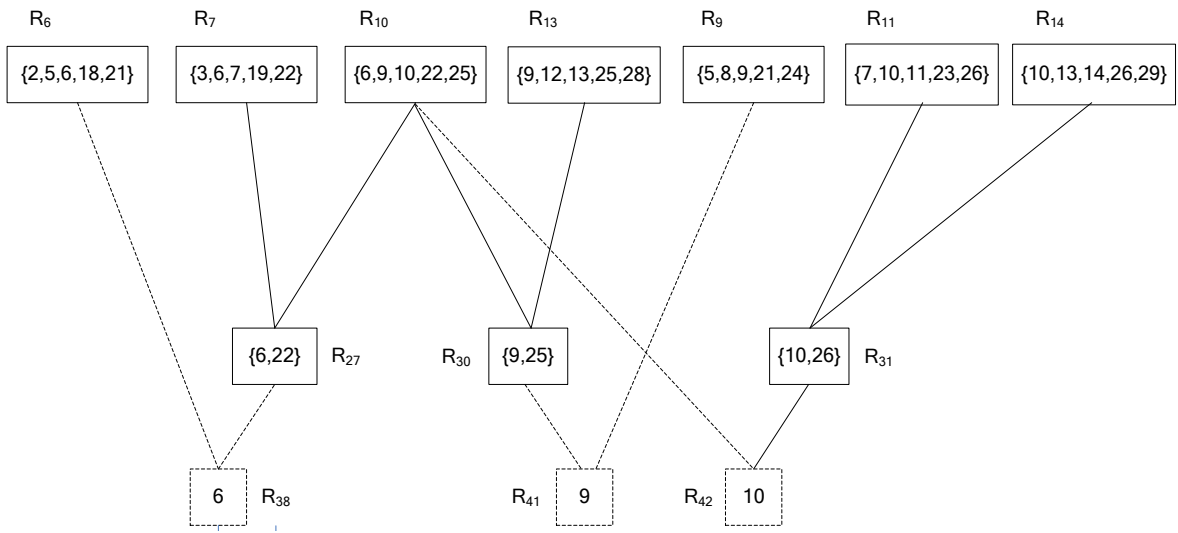

Figure 9. Part of a GBP region graph of a 2D TBCC

Having defined the region graph, we are now ready to apply the GBP algorithm that operates over it. We chose to use the variant named "Two-Way Algorithm" in [12], which is called this way since in this algorithm messages are passed both from parent regions to their children and from children to their parents.

The complexity of the GBP algorithm is governed by the size of its largest region. This region contains $wt(g)$ variable nodes, and therefore has to compute $2^{wt(g)}$ beliefs at each iteration. In contrast with the LBP algorithm, which has a complexity of $O(wt^2(g))$ at most, the complexity is now exponential with $wt(g)$. This is because the GBP algorithm has to compute the joint beliefs for the entire group of variable nodes, the largest of which contains $wt(g)$ nodes. Thus the complexity is $O(2^{wt(g)})$ per information bit. This is still better than the 2D trellis decoder which has complexity that is exponential with the size of the kernels, which is obviously larger than their weight. Of course GBP has also lower complexity than the Viterbi algorithm.

## VIII. RESULTS

We have simulated the performance of the various soft decoding algorithms described above for selected codes over the AWGN channel. The simulated algorithms were:
  a. Viterbi algorithm, applied after reducing the problem to 1D.
  b. The 2D trellis algorithm described above.
  c. Classical loopy belief propagation (LBP), commonly used to decode LDPC codes.
  d. Modified LBP.
  e. GBP with modified graph construction as described above.

We display the results for the sum-product variant of the above algorithms. However, other variants were also tested and display a similar performance. In Table V we summarize the performance of these algorithms for selected codes, and also list the codes minimum distance and inverse encoders (where they exist). The performance curves for these codes are plotted in Figure 10 to Figure 13.

The Viterbi algorithm performs very close to the union bound, and we use it as a reference point from which the degradation of the other, sub-optimal algorithms is measured. It can be seen that the different algorithms have varying degrees of success in decoding, depending on the tested code.

The 2D trellis algorithm, showing a very small degradation of less than 0.5 dB, is very successful for codes with a kernel support of 2x2. For codes with a kernel support of 3x3 it maintains a degradation of around ~1.5dB. Of course, some codes are not decodable at all with the 2D trellis algorithm, as evidenced, for example, by code #3.

The LBP algorithm, which usually performs much worse than the Viterbi, fails almost completely for code #2, though recovers by the modified version. However, for some codes, e.g., code #4, degradation is relatively low.

Finally, the GBP algorithm also has a somewhat inconsistent behavior. While this algorithm rarely fails completely, and usually performs better than the LBP algorithm, it sometimes fails to improve upon it.

In Figure 14 we display performance results of decoding of other codes, and compare them with the best results we got for our 2D TBCCs. The codes chosen for comparison are:

a) Code A, 1D TBCC from section V, using Viterbi algorithm, applied on 3 repetitions of the block to allow tail-biting.
b) LDPC code with rate ½, $k=36$ and $d_v=3$ and $d_c=6$, using LBP.

One can see that the 2D TBCC has slightly better performance than the LDPC code, and is also slightly better than the 1D CC. Note, however, that the potential gain from using a better decoding algorithm is another 1.5dB.

**Table V. Performance of 2D TBCC Decoding Algorithms**

| | Encoder Kernels | | Inverse Kernels | | $d_{min}$ | Viterbi $E_b/N_0$ [dB] for $P_w=10^{-3}$ | Degradation [dB] | | | |
|---|---|---|---|---|---|---|---|---|---|---|
| | $g_1$ | $g_2$ | $q_1$ | $q_2$ | | | 2D-Trellis | LBP | Modified LBP | GBP |
| 1 | 1 1<br>1 0 | 1 1<br>1 1 | 1 | 1 | 6 | 4.25 | 0.36 | 1.96 | 0.20 | 0.40 |
| 2 | 1 1 1<br>1 1 0<br>0 1 0 | 1 1 1<br>1 1 1<br>0 1 0 | 1 | 1 | 12 | 3.35 | 1.74 | 5.30 | 1.40 | 1.67 |
| 3 | 1 1 0<br>1 1 0<br>0 0 1 | 1 0 1<br>1 1 1<br>1 1 1 | N/A | N/A | 12 | 3.42 | N/A | 2.80 | 1.50 | 2.98 |
| 4 | 1 0 0<br>0 0 0<br>0 0 0 | 1 1 1<br>1 1 0<br>1 0 1 | 1 | 0 | 8 | 3.60 | 1.40 | 1.07 | 0.95 | 1.60 |

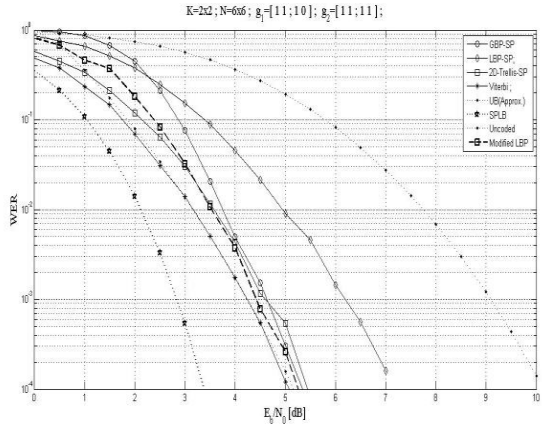

Figure 10. Code #1: Performance curves

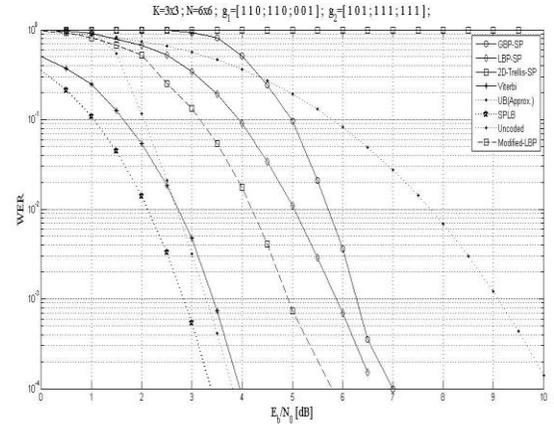

Figure 12. Code #3: Performance curves

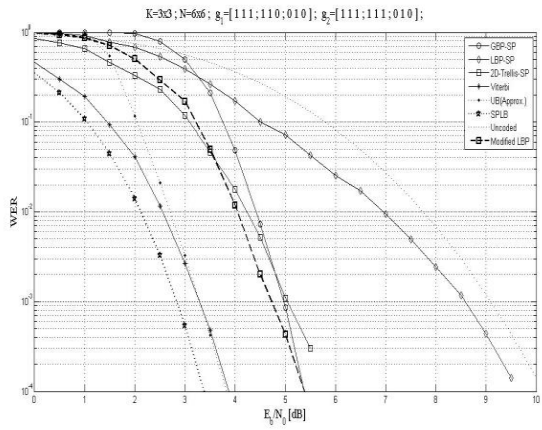

Figure 11. Code #2: Performance curves

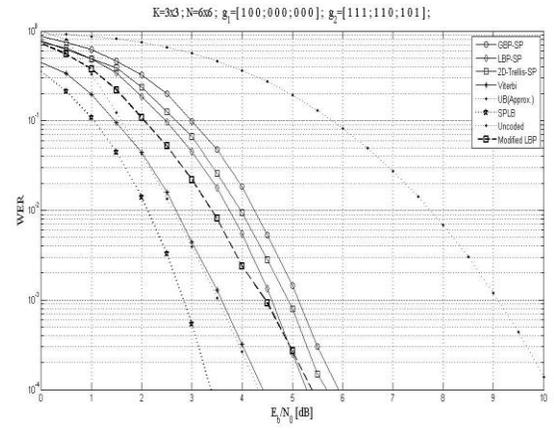

Figure 13. Code #4: Performance curves

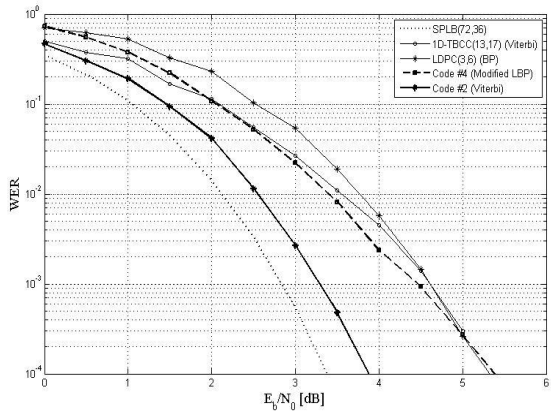

Figure 14. Comparison between different codes

## IX. CONCLUSION

In this paper, we took a systematic approach to analyze 2D TBCCs, in order to address two of the main open problems in the field: establishing the encoding power of these codes, and finding practical decoding algorithms for them. By starting out from basic algebraic properties, we have established basic criteria to test whether a 2D TBCC encoder is non-degenerate, has a parity check matrix or is polynomially-invertible. These criteria give the designer of a 2D TBCC encoder the tools to determine its usefulness and to discard useless encoders. To further refine the set of useful encoders and to find the most powerful encoders within the set, we used the BEAST algorithm (extended to 2D). This enabled us to establish the coding potential of 2D TBCCs, and to conclude that they are at least as powerful, and in some cases better than codes of comparable rate and size (such as 1D CC counterparts, or block codes such as BCH and LDPC).

In order to decode 2D TBCCs, we have applied several sub-optimal decoding algorithms, including conventional loopy belief propagation (LBP), modified LBP, modified generalized belief propagation (GBP), and a completely novel 2D trellis algorithm. For codes with small kernel support, most of these algorithms performed very well, and in some cases very close to the optimal decoding curve. As the kernel support increased, the performance of these sub-optimal algorithms deteriorated, but was still respectable in most cases. The LBP algorithm, which is least suited for these codes, generally displayed the worst performance, while the modified LBP, the GBP and 2D trellis algorithms were more or less consistent in their performance (0.2-0.5 dB away from the optimum for 2x2 kernels, 1-1.5 dB for 3x3 kernels). We conclude that while sub-optimal decoding algorithms can be successfully applied to 2D TBCCs, there is still room to improve the performance of these algorithms, as well as their complexity.

Finally, we note that 2D CCs and TBCCs have yet to find practical application. The authors hope that the progress made in this paper will make 2D CCs and TBCCs more accessible to researchers, and further the study of this little-explored field.

## ACKNOWLEDGEMENTS

The authors wish to thank Dr. Yaron Shani for his help with the algebraic concepts that appear in this paper, for the formulation and proof of the condition for non-degenerate encoders and for formulating the convergence condition of the 2D trellis algorithm. The authors further wish to thank Mr. Nadav Basson for his assistance in improving the paper.

# APPENDIX A

In this appendix we prove that if an ideal $I$ contains a monomial $\mathbf{x}^\alpha = x_1^{\alpha_1} x_2^{\alpha_2} \ldots x_n^{\alpha_n}$ (where $\mathbf{x}^\alpha$ is the monomial of minimal degree in the ideal), then the <u>reduced Gröbner basis</u> (RGB) of the ideal contains the monomial $\mathbf{x}^\alpha$.

Let us first consider some elementary properties of Groebner bases:

1) Given an ideal $I \subset k[\mathbf{x}]$ and a monomial order $>$, the ideal of leading terms is the monomial ideal $\langle LT(I) \rangle \equiv \langle LT(f) \mid f \in I \rangle$.

Explanation: A leading term $LT(f)$ of a polynomial $f$ is the term with the highest degree in $f$. The ideal $\langle LT(I) \rangle$ is the set of the leading terms of all polynomials in $I$.

2) A finite set $G \subseteq I \setminus \{0\}$ is a Gröbner basis for I under $>$ if $\langle LT(g) \mid g \in G \rangle = \langle LT(I) \rangle$.

Explanation: The leading terms of the elements of $G$, span the ideal of leading terms of $I$. Additionally, $G$ is a basis of $I$, which means the elements of $g$ also span $I$.

3) For every $f \in I$, $LT(f)$ divides by $LT(g)$ of one element in $G$.

4) A Gröbner basis G is reduced if for every $g \in G$:
   a) $LT(g)$ divides no term of any element of $G \setminus \{g\}$.
   b) The leading coefficient of $g$, $LC(g) = 1$.

5) Every ideal has a unique reduced Gröbner basis under $>$.

6) If $G$ is a Gröbner basis of $I$, then every $f \in I$ has a unique representation using $G$.

Explanation: if we divide $f$ by $G$ using the long division algorithm then:
   a) the remainder is zero
   b) the quotients are the same regardless of the order of elements in $G$.

We shall now use these properties in the complete proof. Let $G$ be a reduced Gröbner basis of $I$ and let $\mathbf{x}^\alpha$ be the monomial with smallest degree in $I$. Then from (3) it follows that $\mathbf{x}^\alpha$ divides by some the leading term element of $G$, which we denote by $g_1$. From (4a) it follows that $LT(g_1)$ divides no term of $g_i$, where $i \neq 1$, and also that $LC(g_1)=1$. Since $\mathbf{x}^\alpha$ is of minimal degree, it follows that $LT(g_1)= \mathbf{x}^\alpha$.

Now assume that $g_1 \neq \mathbf{x}^\alpha$. Then we can write $g_1= \mathbf{x}^\alpha+\tilde{g}_1$. Since $G$ is a basis of $I$, it follows that there exist $f_1...f_n$ such that

$$\mathbf{x}^\alpha = f_1 g_1 + f_2 g_2 + ... + f_n g_n = f_1(\mathbf{x}^\alpha+\tilde{g}_1) + f_2 g_2 + ... + f_n g_n$$

Switching sides, and adding and subtracting $\tilde{g}_1$

$$0 = f_1(\mathbf{x}^\alpha+\tilde{g}_1) - \mathbf{x}^\alpha - \tilde{g}_1 + f_2 g_2 + ... + f_n g_n + \tilde{g}_1 =$$

And so we get

$$\tilde{g}_1 = (1-f_1)g_1 - f_2 g_2 - ... - f_n g_n$$

So $\tilde{g}_1 \in I$. Therefore from (6), $\tilde{g}_1$ has a unique representation using $G$. Since the leading term of $\tilde{g}_1$ does not divide by the leading term of $g_1$, the coefficient of $g_1$ must be zero and therefore, $f_1=1$. However, since $G$ is an RGB, it follows from (4a) that none of the terms of $\tilde{g}_1$ divide by any of the leading terms of $g_2,...,g_n$. Therefore $f_2=...=f_n=0$. Thus, $\tilde{g}_1=0$, and $g_1= \mathbf{x}^\alpha$ □